\documentstyle[11pt,amsfonts]{article}

\newcommand{\be}{\begin{equation}}
\newcommand{\ee}{\end{equation}}
\newcommand{\bea}{\begin{array}}
\newcommand{\ea}{\end{array}}
\newcommand{\beqa}{\begin{eqnarray}}
\newcommand{\eeqa}{\end{eqnarray}}
\newcommand{\bean}{\begin{eqnarray*}}
\newcommand{\eean}{\end{eqnarray*}}

\def\starwedge{\stackrel{*}{\wedge}}

\def\BI{{\rm 1\!l}}

\def\up#1{\leavevmode \raise.16ex\hbox{#1}}

\newcommand{\gapproxeq}{\lower
 .7ex\hbox{$\;\stackrel{\textstyle >}{\sim}\;$}}
\newcommand{\lapproxeq}{\lower .7ex\hbox{$\;\stackrel
{\textstyle <}{\sim}\;$}}

\newcounter{appendice}

\def\thebibliography#1{{\bf REFERENCES\markboth
 {REFERENCES}{REFERENCES}}\list
 {[\arabic{enumi}]}{\settowidth\labelwidth{[#1]}\leftmargin\labelwidth
 \advance\leftmargin\labelsep
 \usecounter{enumi}}
 \def\newblock{\hskip .11em plus .33em minus -.07em}
 \sloppy
 \sfcode`\.=1000\relax}

\begin{document}
\begin{flushright}
SU 4252-818\\
\end{flushright}


\centerline{ \LARGE  Noncommutative $AdS^3$ with }

\vskip .5cm
\centerline{\LARGE  Quantized
  Cosmological Constant}

\vskip 2cm

\centerline{ {\sc    A. Pinzul$^{a}$ and A. Stern$^{b}$ }  }

\vskip 1cm
\begin{center}
{\it a)  Department of Physics, Syracuse University,\\ Syracuse,
New York 13244-1130,  USA\\}
{\it b) Department of Physics, University of Alabama,\\
Tuscaloosa, Alabama 35487, USA}

\end{center}

\vskip 2cm

\vspace*{5mm}

\normalsize
\centerline{\bf ABSTRACT}
We examine a recent deformation of  three-dimensional anti-deSitter
gravity based on  noncommutative Chern-Simons theory with gauge group
$U(1,1)\times U(1,1)$.  In addition to  a noncommutative analogue of
$3D$ gravity, the theory contains two addition gauge fields which decouple in
the commutative limit.  It is well known that the level is quantized
in noncommutative  Chern-Simons theory.  Here it implies that the
cosmological constant goes like minus one over an integer-squared.  We
construct the noncommutative  $AdS^3$ vacuum by applying a
Seiberg-Witten map from the commutative case.  The procedure is
repeated for the case of a conical space resulting from a massive
spinning particle.

\vspace{2cm}

{\it Dedicated to Rafael Sorkin in his 60th year.}

\vspace*{5mm}

\newpage

\section{Introduction}

\setcounter{equation}{0}

There is current interest in writing down  noncommutative deformations
of Einstein gravity as they may serve to  model  quantum
gravity.\cite{Behr:2003qc},\cite{Banados:2001xw},\cite{Cacciatori:2002gq},\cite{Bimonte:1997dw},\cite{Chamseddine:2003cw}
Some of the proposals  are technically involved and their physical
implications are not evident.  It is therefore  helpful to examine
simple models.     One such example was proposed by Banados,
et. al. \cite{Banados:2001xw} which is applicable to  three
dimensional gravity and assumes  a constant noncommutativity.  That
approach is a deformation of    the
Chern-Simons  description of $3D$ gravity.\cite{Achucarro:1987vz},\cite{Witten:1988hc} The
gauge group in the commutative Chern-Simons theory can be
$ISO(2,1)$, $SO(3,1)$ or $SO(2,2)$,
corresponding  to flat, deSitter ($dS$) or anti-deSitter ($AdS$) space,
respectively.    The deformation to the corresponding   noncommutative
theory has been given using gauge groups $GL(2,C)$ and $U(1,1)\times
 U(1,1)$, which introduces two additional gauge fields that were
not present in the commutative theory.\footnote{Noncommutative Chern-Simons gravity
  has also been written down using q-deformed gauge groups.\cite{Bimonte:1997dw}} They decouple in the
commutative limit, giving rise to two $U(1)$  gauge fields.  One
feature of noncommutative Chern-Simons theory is the quantization of
the overall constant in the action, or level,  which, unlike in
commutative Chern-Simons theory,
occurs even when the gauge group is $U(1)$.\cite{Nair:2001rt},\cite{Bak:2001ze}   Applying this to
Chern-Simons gravity leads to the following quantization of the
cosmological constant $\Lambda$ times the gravitational coupling $G$
squared
in the noncommutative $AdS^3$:
 \be \Lambda G^2 = - \frac 1{16  k^2}
\label{cscntqntzn}\;, \ee where $k$ is the integer level.  A  similar result was found some time
ago following a very different approach.\cite{Moss:1991as}  Rather than involving
noncommutativity it relied on gravitational instantons and was
applicable in four dimensions.

    Chern-Simons theory is trivial  in the absence of
 sources and  written on topologically trivial   manifold;
 i.e. Chern-Simons theory on
the  plane contains no dynamical degrees of freedom; nor does  Chern-Simons theory on
the noncommutative plane.  Moreover, one can
Seiberg-Witten   map  free commutative  Chern-Simons theory to
 the corresponding
noncommutative  Chern-Simons theory.\cite{Grandi:2000av} So in Chern-Simons gravity,
 the $AdS^3$ vacuum solution is mapped  to a solution of the
free noncommutative gravity  equations.  We use this procedure to obtain the noncommutative
 $AdS^3$ vacuum solution at leading order in the noncommutativity
 parameter.  At this order, the spin connections and triads are
 undeformed, but the two additional gauge fields are nonvanishing on the
 spatial domain.

  Sources can be introduced in commutative gauge theory by removing points from the space.   In the case of $3D$
gravity, such sources represent conical singularities, which upon
embedding in $4D$ space-time can
serve to
model cosmic strings.  The
noncommutative version of this system may give insight into the
 quantum theory.  Point sources have been introduced previously in $3D$
 noncommutative
 gravity in the context of matrix algebra.\cite{Shiraishi:2002qu},\cite{Valtancoli:2005js}   Our
 approach will be to Seiberg-Witten  map the conical solutions  of the commutative theory
 to the noncommutative theory  and  obtain the leading corrections to the
 commutative theory.
 The result shows that
  a cloud of    flux associated with the two additional
gauge fields  surrounds the conical
singularity in the noncommutative theory.   Moreover, if $U(1)$ fluxes are present in  the
 commutative theory, the Seiberg-Witten map produces  radially-dependent corrections to
 the effective noncommutative mass and spin.

After giving the argument for the cosmological constant quantization
in section 2, the  field equations for
  commutative and noncommutative Chern-Simons gravity
are reviewed in section 3.  We write down solutions of the commutative
equations in section 4.  They correspond to the $AdS^3$ vacuum and the conical space resulting from a massive
spinning particle.   In section 5 we apply the Seiberg-Witten map  to
obtain  solutions of
the noncommutative theory at lowest order in the noncommutativity
  parameter.

\section{Quantization of the cosmological constant}
\setcounter{equation}{0}

Here  we review commutative Chern-Simons gravity\cite{Witten:1988hc},
the noncommutative version as given in \cite{Cacciatori:2002gq}, and
finally level quantization\cite{Nair:2001rt},\cite{Bak:2001ze}, and
show that this leads to (\ref{cscntqntzn}).

\subsection{Commutative Chern-Simons gravity}

Up to boundary terms, the Einstein-Hilbert action with negative
cosmological constant $\Lambda$ is the difference of two $SU(1,1)$ Chern-Simons
actions
\be {\cal   S}({\cal A}^+,{\cal A}^-) = {\cal  S}_+({\cal A}^+) - {\cal   S}_-({\cal A}^-) \;,\label{cmtvactn}\ee
\be {\cal   S}_\pm({\cal A}^\pm) = \beta \int {\rm tr} \biggl(  {\cal A}^\pm \wedge d{ \cal A}^\pm
+\frac 23  {\cal A}^\pm \wedge{\cal  A}^\pm \wedge {\cal   A}^\pm \biggr)\;\ee
 $\beta$ is the dimensionless coupling constant which expressed in
terms of the gravitational coupling $G$ and $\ell =1/ \sqrt{-\Lambda}$ is
\be \beta = \frac {\ell}{16\pi G}\;,\label{btntmsflmbd} \ee
and the $SU(1,1)$ connection one forms ${\cal   A}^\pm$, which gauge
transform according to \be  {\cal   A}^\pm \rightarrow  {\cal  A}'^\pm = g_\pm^{-1}  {\cal  A}^\pm g_\pm + g_\pm^{-1}
d g_\pm \;,\label{ggtrns} \ee $ g_\pm$ taking value in $SU(1,1)$.
${\cal   A}^\pm$ are written in terms
of triads $e^a$, $a=0,1,2$, and spin connection one forms $\omega^a$ according to
 \be{\cal   A}^\pm ={\cal   A}^{a\pm} \tau_a\;,\qquad {\cal  A}^{a\pm} = \omega^a \pm
 e^a/\ell\;\label{su11pot}\ee   $\tau_a$
are  $SU(1,1)$  generators  \be [\tau_a,\tau_b]
= \epsilon_{abc} \tau^c \;, \ee  indices $a,b,c,...=0,1,2$ are raised and lowered with the Lorentzian metric $\eta
= {\rm diag}(-1,1,1)$, and we take $\epsilon^{012}=-\epsilon_{012}=1$.
 We choose the defining representation
\be \tau_0 = \frac i 2 \sigma_3\quad \tau_1 = \frac 12 \sigma_1\quad
\tau_2 = \frac 1 2 \sigma_2\;, \ee $\sigma_a $ being the three Pauli
matrices.
Then $\tau_a\tau_b
=\frac 12 \epsilon_{abc} \tau^c +\frac 14 \eta_{ab}\BI $ and
Tr$\tau_a\tau_b =\frac 12 \eta_{ab}$.  Substituting into
(\ref{cmtvactn}) gives
\be {\cal   S}({\cal  A}^+,{\cal  A}^-) =  \frac {1}{8\pi G} \int\biggl( e^a \wedge R_a  +\frac
1{6\ell^2} \epsilon_{abc} e^a\wedge e^b\wedge  e^c
\biggr)\;,\label{actfreaom} \ee up to boundary terms,
where \be R^a =d\omega^a +\frac 12 \epsilon^{abc}
\omega_b\wedge\omega_c \; \ee is the spin curvature two-form.

  Following  \cite{Cacciatori:2002gq} the  gauge group has to be
enlarged in the noncommutative theory  to    $U(1,1)\times U(1,1)
$.\footnote{An alternative approach would be to construct the
  noncommutative theory from the commutative  $SU(1,1)\times SU(1,1)
$ theory using the Seiberg-Witten map as described in
\cite{Jurco:2001rq}.  This would then not require the enlargement of
the gauge group.  However, a  nonperturbative formulation of this approach
may not be simple.}
The resulting  gauge theory then contains two additional   potentials one forms $b^\pm$,
which were not present in the above system.
  Here one  replaces  $SU(1,1)$
connection one forms $ {\cal  A}^\pm$ in (\ref{su11pot}) with  $U(1,1)$
connection one forms
\be {\cal   A}^{a\pm} \tau_a +  b^\pm \tau_3\;,\qquad {\cal  A}^{a\pm} = \omega^a \pm
 e^a/\ell \label{gldfofapm}\;,\ee  where
  $\tau_a$, $a=0,1,2$, along with the central element
$$ \tau_3 = \frac i2\BI \;,$$ span the   $U(1,1)$
generators, and one enlarges the set of gauge transformations  $\{ g_\pm\}$ in (\ref{ggtrns}) to $U(1,1)$.   Substituting into (\ref{cmtvactn}) now gives
(\ref{actfreaom}) plus
\be -\frac \beta 2 \int (b^+ d b^+ -b^- d b^-)\;. \label{trvlctvcs} \ee  Thus $b^\pm$
 represent two noninteracting Abelian gauge
potentials in the commutative theory having trivial dynamics.  This is not the case in the
noncommutative analogue of the theory.

\subsection{Noncommutative Chern-Simons gravity}

 The noncommutative analogue of Chern-Simons
gravity was  given previously in \cite{Banados:2001xw},
\cite{Cacciatori:2002gq}.   Noncommutativity is
realized by replacing the commutative product on the plane by the   Groenewold-Moyal star product
\be \star = \exp\;\biggl\{ \frac {i}2 \theta^{\alpha\beta}\overleftarrow{
  \partial_\alpha}\;\overrightarrow{ \partial_\beta} \biggr\}  \ee
where $\theta^{\alpha\beta}=-\theta^{\beta\alpha}$ is  the
noncommutative matrix and
  $\overleftarrow{
  \partial_\alpha}$ and $\overrightarrow{ \partial_\beta}$ are left and right
derivatives on the commuting plane, respectively.  The noncommutative
analogue of the action (\ref{cmtvactn}) is
\be  \hat {\cal   S}(\hat {\cal  A}^+,\hat {\cal   A}^-) = \hat{\cal   S}_+(\hat {\cal  A}^+) - \hat {\cal   S}_-(\hat{\cal   A}^-) \;,\label{ncmtvactn}\ee
\be \hat {\cal   S}_\pm(\hat{\cal   A}^\pm) = \beta \int {\rm tr} \biggl( \hat {\cal   A}^\pm
\starwedge d \hat{\cal   A}^\pm
+\frac23 \hat {\cal   A}^\pm \starwedge \hat{\cal   A}^\pm \starwedge \hat{\cal   A}^\pm
\biggr)\;,\label{paactns}\ee
 where $\starwedge$
means one takes a  Groenewold-Moyal star product between components of the
forms and $ {\cal   A^\pm}$ are  $U(1,1)$
connection one forms, which gauge
transform according to
\be \hat  {\cal  A}^\pm \rightarrow \hat {\cal   A}'^\pm = \hat g_\pm^{-1}\star \hat {\cal   A}^\pm\star\hat g_\pm +\hat g_\pm^{-1}\star
d\hat  g_\pm \;,\label{ncggtrns} \ee $\hat g_\pm$ taking values in
noncommutative $U(1,1)$, with $\hat g_\pm^{-1}\star
\hat  g_\pm = 1 .$
 We expand $\hat{\cal   A}^\pm$ in terms of connection components
$\hat e^a$, $\hat \omega^a$ and $\hat b^\pm$, which respectively are the noncommutative analogues of triads $e^a$,  spin
connection one forms $\omega^a$ and $U(1)$ one forms $b^\pm$
\be \hat {\cal   A}^\pm = \hat {\cal   A}^{a\pm} \tau_a +  \hat b^\pm  \tau_3\;,\qquad\hat {\cal   A}^{a\pm} = \hat\omega^a \pm
\hat e^a/\ell\;. \label{ncgldfofapm}\ee  Upon substituting into
(\ref{ncmtvactn}) one gets
\beqa  & &\hat {\cal   S}( {\cal A }^+,\hat  {\cal  A}^-) \;=\;\frac {1}{8\pi G} \int\biggl(\hat e^a \starwedge\hat R_a  +\frac
1{6\ell^2} \epsilon_{abc} \hat e^a\starwedge \hat e^b\starwedge  \hat e^c
\biggr)\cr & &\cr & &+\; \frac \beta 2 \int \biggl( \hat b^- \starwedge d \hat
b^- +\frac i3 \hat b^-\starwedge \hat b^-\starwedge \hat b^-
\biggr)\;-
\; \frac \beta 2 \int \biggl( \hat b^+ \starwedge d \hat
b^+ +\frac i3 \hat b^+\starwedge \hat b^+\starwedge \hat b^+
\biggr)\\ & &\cr & &+\; \frac {i\beta} 2 \int (\hat b^+ -\hat b^-)
\starwedge\biggl( \hat \omega^a\starwedge \hat\omega_a + \frac
1{\ell^2}\hat e^a\starwedge\hat e_a \biggr)+\; \frac {i\beta} {2\ell} \int (\hat b^+ +\hat b^-)
\starwedge\biggl( \hat \omega^a\starwedge \hat e_a +\hat e^a\starwedge \hat \omega_a\biggr)\nonumber
\;,\label{ncactfreaom} \eeqa  up to boundary terms,
where \be \hat R^a= d\hat \omega^a +\frac 12 \epsilon^{abc}
\hat  \omega_b\starwedge\hat\omega_c \; \ee is the noncommutative analogue
of the spin curvature two-form.  The analogous two dimensional system
was given in \cite{Cacciatori:2002ib}.

\subsection{Level quantization}

We finally recall the  work of \cite{Nair:2001rt},\cite{Bak:2001ze}.
Here it is more convenient to use the infinite dimensional matrix
realization of Chern-Simons theory.   We assume that time $t$ is
commutative, while the
space coordinates $\hat x_1$ and $\hat x_2$ generate a Moyal plane
\be [\hat x_1,\hat x_2] = i \theta \;.\label{cmtrxoxt}\ee
They, along with the $U(1,1)$ generators $\tau_a$, act on   a  Hilbert space $H$, with a basis $|n,s>$, $n=0,1,2,...$,
$s=1,2$, according to
  \beqa (\hat x_1 +i\hat x_2) |n,s> &=&\sqrt{2\theta n}\; |n-1,
s>\\
& &\cr \tau_a |n,r> &=& (\tau_a)_{rs}\;|n,s>
\;. \eeqa
  Next introduce  potentials $\hat
A^\pm_\mu$, $\mu=0,1,2$, on the noncommutative plane  $\times\;
{\mathbb{R}}^1$, which are valued in the Lie-algebra of $U(1,1)$,
meaning that
\be  \hat
A^\pm_\mu =\hat
A^{a\pm }_\mu(t,\hat x_1,\hat x_2)\; \tau_a\;.\ee
 The map from $\hat
A^\pm_\mu$ to functions $\hat{ \cal A}^\pm_\mu$ on ${\mathbb{R}}^3$ is the symbol map, the latter denoting the space-time
components of the one form $\hat {\cal A}^\pm$. Gauge
transform are generated by   unitary operators $\hat U_\pm$
\beqa  \hat  {  A}^\pm_0& \rightarrow & \hat {  A}'^\pm_0 = \hat U_\pm^{-1} \hat {   A}^\pm_0\hat U_\pm +\hat U_\pm^{-1}\partial_0
\hat  U_\pm \cr & &\cr
 \hat  {  A}^\pm_i& \rightarrow & \hat {  A}'^\pm_i = \hat U_\pm^{-1}
 \hat {   A}^\pm_i\hat U_\pm +\frac i\theta\epsilon_{ij}\;\hat U_\pm^{-1}[\hat x_j,
\hat  U_\pm] \;,\qquad i,j=1,2 \;. \eeqa   To obtain the operator
analogue  of the action (\ref{paactns}) we replace symbols by the
corresponding operators and  the space integral $\int d^2x$ by
$2\pi\theta\;{\rm Tr}$, where
 the trace is over all of $H$.   Up to boundary terms,
one gets
\be \hat S_\pm(\hat
A^\pm) =\frac{\theta k}2 \int dt\; {\rm Tr} \biggl( -\epsilon_{ij} \hat A^\pm_i \partial_0 \hat A^\pm_j
+ 2 \hat A^\pm_0 \hat F^\pm_{12} \biggr)\;,\label{offrtactn}\ee
where \be k = 4\pi \beta \label{lvl}\ee
is the level and
\be \hat F^\pm_{12}= \frac i\theta [\hat x_2, \hat A^\pm_2]   +\frac
i\theta [\hat x_1, \hat A^\pm_1] +[  \hat A^\pm_1,\hat A^\pm_2]\;,
\ee
which transform  covariantly $\hat F^\pm_{12}\rightarrow \hat F^{'\pm}_{12}=
 \hat U_\pm^{-1}\hat F^\pm_{12} \hat U_\pm$.  Another set of
 covariant operators is
\be \hat X^\pm_i=\frac i\theta\epsilon_{ij}\;\hat x_j   +\hat
A^\pm_i\;, \ee in terms of which the action (\ref{offrtactn}) becomes
\be \hat S_\pm(\hat
A^\pm) =k \int dt\; {\rm Tr} \biggl( -\frac{\theta }2\epsilon_{ij} \hat X^\pm_i D_0 \hat X^\pm_j
+ i \hat A^\pm_0  \biggr)\;,\ee
where $D_0\hat X^\pm_i = \partial_0\hat X^\pm_i + [\hat A^\pm_0,\hat
X^\pm_i]$ is covariant under gauge transformations and  we have dropped total time derivatives in the integrand.
The first term in the trace is invariant, while the second is not.
As a result, a  gauge transformations induces the following change in the action is
\be \Delta\hat S_\pm = i k \int dt\; {\rm Tr}\hat U_\pm^{-1}\partial_0
\hat  U_\pm\;.\label{chngnactn} \ee
Following \cite{Nair:2001rt},\cite{Bak:2001ze}, one imposes the boundary
condition $\hat U_\pm|n,s>=|n,s>$ as $n\rightarrow\infty$  on the
gauge transformations, effectively making $\hat U_\pm$  elements of
$U(N,N)$, for some large $N$.
(\ref{chngnactn}) is then proportional to the winding number of
noncontractable loops in  $U(N,N)$ parametrized by the time. Writing $\hat U_\pm(t)=
e^{i\alpha(t)} \hat V(t)$, where $\hat V(t)\in SU(N,N)$,  a typical  noncontractable loop is given
by  \be
\alpha(-\infty)=0\;,\quad V(-\infty)=\BI \;,\qquad\quad \alpha(\infty)=-\frac{\pi n}N\;,\quad V(\infty)=
e^{\frac {i\pi n}N}\BI, \ee where $n$ is an integer$\ne 0$.
Substituting into (\ref{chngnactn}) gives $\Delta\hat S_\pm = 2\pi k
n$.  From the singlevaluedness of $\exp{i\hat S_\pm}$, it then follows
that $k$ is an integer.  The quantization condition (\ref{cscntqntzn})
results from (\ref{lvl}).  Note that the $U(1)$ degrees of freedom
play a central role in this discussion.

\section{Field equations}
\setcounter{equation}{0}

The field equations following from the commutative Chern-Simons action (\ref{actfreaom})  are
 \beqa
R^a  +\frac 1 {2\ell^2}\epsilon^{abc} e_b\wedge  e_c &=&0 \cr & &\cr   T^a\equiv de^a + \epsilon^{abc} \omega _b\wedge e_c & =& 0\;, \label{zceo} \eeqa
along with \be d b^\pm = 0 \label{trvldynmcsfrb}\;,  \ee  for the
 $U(1)$ fields.
(\ref{zceo}) states that  the  $SU(1,1)\times SU(1,1)$ curvature vanishes
 \be  {\cal   F}^\pm \equiv d {\cal   A}^{\pm}
+  {\cal  A}^\pm\wedge  {\cal  A}^\pm =0\;. \label{cseqswosc}\ee  The solutions, at least
locally, can be expressed as pure gauges \be  {\cal  A}^\pm = g_\pm^{-1}
d g_\pm \;, \label{prgg}\ee where $g_\pm$ are  $SU(1,1)$ group elements. Group
elements $g$ in the defining representation of  $SU(1,1)$ satisfy
\be \BI= \sigma_3 g^\dagger \sigma_3 g =  -4 \tau_0  g^\dagger \tau_0g
\;, \label{dfcdnge}\ee along with det$g=1$.

The noncommutative field equations following from (\ref{ncactfreaom})  are\cite{Cacciatori:2002gq}
 \beqa  \hat R^a  +\frac 1 {2\ell^2}\epsilon^{abc}\hat e_b\starwedge  \hat e_c &=&-\frac{i} 4 \bigg[
\hat b^-,\hat \omega^a -\frac 1\ell\hat e^a\biggr]_{\star +}-\frac{i} 4 \bigg[
\hat b^+,\hat \omega^a +\frac 1\ell\hat e^a\biggr]_{\star +}
\cr & &\cr
\hat T^a\equiv d\hat e^a +\frac 12 \epsilon^{abc} (\hat \omega _b\starwedge
\hat e_c+\hat e _b\starwedge \hat \omega_c) & =&\;\frac{i\ell} 4 \bigg[
\hat b^-,\hat \omega^a -\frac 1\ell\hat e^a\biggr]_{\star +}-\frac{i\ell} 4 \bigg[
\hat b^+,\hat \omega^a +\frac 1\ell\hat e^a\biggr]_{\star +}
\cr & &\cr   \hat B^\pm =d \hat b ^\pm  +\frac i2 \hat b^\pm\starwedge \hat
 b^\pm &=& \frac i2 \biggl(\hat \omega^a \pm\frac 1\ell\hat e^a\biggr) \starwedge \biggl(\hat \omega_a \pm\frac 1\ell\hat e_a\biggr)
\;, \label{nczceo} \eeqa
where $[\hat a,\hat b]_{\star +}=\hat a\starwedge\hat b+\hat b\starwedge\hat a
$.  The right hand side vanishes in the commutative limit and one
recovers (\ref{zceo}) and (\ref{trvldynmcsfrb}).
(\ref{nczceo}) states that  the  noncommutative  $U(1,1)\times U(1,1)$ curvature vanishes
 \be \hat {\cal   F}^\pm \equiv d \hat {\cal   A}^{\pm}
+ \hat {\cal   A}^\pm\starwedge  \hat {\cal   A}^\pm =0 \label{nccseqswosc}\ee
The solutions, at least
locally, can again be expressed as pure gauges
 \be\hat  {\cal  A}^\pm =\hat g_{\pm}^{-1}\star
d \hat g_{\pm }\;, \ee     where $\hat g=\hat g_\pm$ are  noncommutative $U(1,1)$ group elements,
which satisfy
\be \BI= \sigma_3 \hat g^\dagger \sigma_3\star\hat g =  -4 \tau_0\hat
g^\dagger \tau_0\star \hat g
\;, \label{dfcdngestr}\ee now
 with no restriction on the determinant, and $\hat g^{-1}\star
 \hat g= \BI$.  Hence  $\hat g^{-1}= -4 \tau_0\hat
g^\dagger \tau_0$.

In the next two sections we write down solutions of the commutative
equations  (\ref{zceo}) and Seiberg-Witten map them to solutions of
the noncommutative equations (\ref{nczceo}).

\section{Solutions to the  commutative theory}
\setcounter{equation}{0}

We first write down the $AdS^3$ vacuum solution and then the  conical
solution resulting from a massive spinning source.  In both cases,
$b^\pm=0$.  This restriction is dropped in the final example, as we allow the source to have nonvanishing $U(1)$ fluxes.

\subsection{The $AdS^3$ vacuum}

Start with the  $AdS^3$ vacuum  metric
\be ds^2 = \frac{-(1+\frac{ r^2}{4\ell^2} )^2 dt^2+ dx_1^2+dx_2^2}{(1-\frac{ r^2}{4\ell^2}
  )^2}\;,\label{adsvcmmtrc}\ee where $r^2=x_1^2+x_2^2$ .  The
corresponding triads and spin connections can be given by
\be \matrix{ e^0 = \frac{1+\frac{ r^2}{4\ell^2} }{1-\frac{ r^2}{4\ell^2 } }\;
dt & &  e^i = \frac{ dx_i
}{1-\frac{ r^2}{4\ell^2 } }  \cr & & \cr
 \omega^0 = -\frac{1 }{2\ell^2 }\;\frac{ \epsilon_{ij} x_i dx_j
 }{1-\frac{ r^2}{4\ell^2 } } & & \omega^i =\frac 1{\ell^2 }\; \frac{ \epsilon_{ij}
   x_j dt  }{1-\frac{ r^2}{4\ell^2 }
}} \;,\label{vcsln} \ee where $i,j,..$ denote the spatial components,
$i,j,.. =1,2$, and $\epsilon_{12}=-\epsilon_{21}=1$.  They satisfy
(\ref{zceo}), or equivalently  (\ref{cseqswosc}), and so solutions can
be written as
pure gauges (\ref{prgg}). The  $SU(1,1)$
group elements $g_\pm$ appearing in   (\ref{prgg}) which give (\ref{vcsln}) are
\be g_\pm = \frac 1 { \sqrt{1-\frac
    { r^2}{ 4\ell^2}}} \;\pmatrix{ \exp{ \frac{\pm it }{2\ell}}
&  \pm\frac{1 }{2\ell}\;\bar z\;\exp{ \frac{\pm it
  }{2\ell}}\cr\pm \frac{1}{2\ell}\;z\;\exp{ \frac{\mp it
  }{2\ell}} &  \exp{ \frac{\mp it }{2\ell}}\cr
}\label{grplmnts} \;, \ee where $z=x_1+i x_2$ and $\bar z=x_1-i x_2$.
$ g_\pm$ represents the gauge transformation from the zero
connection to the vacuum solution (\ref{vcsln}).

\subsection{Conical solutions on $AdS^3$}

The dynamics of a massive spinning test particle in the $AdS^3$ vacuum
was examined in \cite{Skagerstam:1989ti}.  Here we instead regard the massive spinning
particle as a source.  For a source at the origin the field equations are \beqa
R^a  +\frac 1 {2\ell^2}\epsilon^{abc} e_b\wedge  e_c &=&2\pi m\;
 \delta^a_0 \delta(x_1)\delta(x_2)d^2x \cr & &\cr   T^a\equiv de^a + \epsilon^{abc} \omega _b\wedge e_c & =& 2\pi s\;
 \delta^a_0 \delta(x_1)\delta(x_2)d^2x\;,  \label{zceows} \eeqa or
\be  {\cal   F}^\pm \equiv d  {\cal  A}^{\pm}
+  {\cal  A}^\pm\wedge  {\cal  A}^\pm =2\pi\tau_0 \biggl( m\pm \frac
s\ell\biggr)\delta(x_1)\delta(x_2)d^2x  \;. \ee
In the
 gauge where only the time components  are
nonvanishing  the solutions are
\be e^0 = s\; d\phi \qquad \omega^0 = m\; d\phi\qquad e^i=\omega^i=0\;,\qquad r>0 \;,\label{snglrmtrc}\ee or
\be   {\cal  A}^\pm  =\biggl( m\pm \frac s\ell\biggr)\;d\phi\; \tau_0\;,\qquad r>0\label{snglrptnl}
\ee  where $\phi $ is
the polar coordinate.
This connection can be expressed as a pure gauge, but only if  one
uses  singular group elements
\be  {\cal  A}^\pm  =e^{- ( m\pm \frac s\ell)\tau_0\phi}\; d  e^{
  ( m\pm \frac s\ell)\tau_0\phi}\;. \ee
 The metric associated with (\ref{snglrmtrc})
is of course singular.  To obtain a nonsingular metric one can
perform a gauge transformation (\ref{ggtrns})  by $g_\pm$ given in (\ref{grplmnts}).
  We then
get the following resulting
 triads and spin connections, which we now denote respectively by    $e'^a$  and
$\omega'^a$:
\be \matrix{ e'^0 = \frac{1+\frac{ r^2}{4\ell^2} }{1-\frac{ r^2}{4\ell^2} }\;
(dt + s\; d\phi) & &  e'^i = \frac{(1-m)\delta_{ij} \; +\; m \hat x_i\hat x_j
}{1-\frac{ r^2}{4\ell^2} }\;dx_j  \cr & & \cr
 \omega'^0 = \frac{m(1+\frac{ r^2}{4\ell^2})\;-\;\frac{ r^2 }{2\ell^2}
 }{1-\frac{ r^2}{4\ell^2} }\;d\phi & &\quad \omega'^i = \frac 1{\ell^2}\;\frac{ \epsilon_{ij}
   x_j dt\; +\; s (\hat x_i\hat x_j-\delta_{ij}) dx_j   }{1-\frac{ r^2}{4\ell^2}
}}\;,\qquad r>0 \;,\label{scsln} \ee where $\hat x_i = x_i/r$.
  $e'^a$  and
$\omega'^a$ in  (\ref{scsln}) are also solutions to (\ref{zceows}).
This is because the source term is invariant under gauge
transformations by $g_\pm$,  as $g_\pm$ at the origin is a rotation
about $\tau_0$, $g_\pm|_{r=0} = \exp( \mp  t
\;\tau_0/\ell)$.  The triads in (\ref{scsln}) are associated with the
 metric
\be ds^2 = \frac{-(1+\frac{ r^2}{4\ell^2} )^2\;(  dt + s \; d\phi)^2\; +\;
  dr^2\;+\; (1-m)^2 \; r^2 d\phi^2}{(1-\frac{ r^2}{4\ell^2}
  )^2}\;.\label{cmtvsnglrmtrc}\ee
In the limit $\Lambda\rightarrow 0$, the connections (\ref{scsln})
reduce to those in \cite{deSousaGerbert:1990yp}, and  (\ref{cmtvsnglrmtrc}) agrees with the  familiar  metric of \cite{Deser:1983tn} upon a rescaling of the
radial coordinate.

\subsection{$U(1)$ Fluxes}

The $U(1)$ fields were assumed to vanish in the previous two
examples.  More generally,   we can  assign $U(1)$ fluxes $\Phi_\pm$  to the
source, giving rise to the potentials
\be  b^\pm  =\Phi_\pm \;d\phi\; \;,\qquad r>0\;. \label{nvngflx}
\ee
The total connection one forms are then
\be  {\cal   A}^\pm  =\biggl\{\biggl( m\pm \frac s\ell\biggr) \tau_0+\Phi_\pm \tau_3 \biggr\} \;d\phi\;\;,\qquad r>0\;.\label{snglrptnlwnzb}
\ee
  The $U(1)$
potentials are unchanged by the gauge transformation (\ref{ggtrns}),
and thus (\ref{scsln}) plus (\ref{nvngflx}) define the general case.

\section{Seiberg-Witten map to noncommutative Chern-Simons gravity}
\setcounter{equation}{0}

It was shown in \cite{Grandi:2000av} that  free noncommutative Chern-Simons theory is
identical to its commutative counterpart, implying that all solutions of
the free commutative theory can be Seiberg-Witten mapped to  all solutions of
the free noncommutative theory.
Up to first order in
$\theta^{\alpha\beta}$, the Seiberg-Witten map is given
by\cite{Jurco:2001rq}
\be \hat {\cal   A}^\pm_\mu =   \hat {\cal   A}^\pm_\mu[ {\cal  A}^\pm] = {\cal  A}^\pm_\mu - \frac
i4\theta^{\alpha\beta}   [  {\cal  A}^\pm_\alpha , \partial_\beta  {\cal   A}^\pm_\mu+
 {\cal  F}^\pm_{\beta\mu}]_+\;,\label{swm} \ee where $\mu,\nu,.. $ are the space-time
components of the forms and $[\;,\;]_+$ denotes an anticommutator.   We
shall assume that there is no time-space noncommutativity, i.e. $
\theta^{t1}=\theta^{t2}=0,\;\theta^{12}=\theta$.

Below we  Seiberg-Witten map the commutative solutions of the previous
section to their
noncommutative counterpart.   In the first two examples, $\hat b^\pm$ vanishes at lowest
order in $\theta$, but picks up a first order contribution from the
Seiberg-Witten map.  In fact, there are no other first order
contributions, i.e.  there are no corrections at this order  to the gravitational
fields, $\hat e^a= e^a +{\cal O}(\theta^2)$ and $\hat \omega^a=  \omega^a+{\cal O}(\theta^2)$.  The
noncommutative analogue of the  $AdS^3$ vacuum  is a solution of the free
noncommutative theory.

\subsection{Noncommutative $AdS^3$ vacuum}

 If we substitute   the $AdS^3$ vacuum solution
(\ref{vcsln}) into (\ref{swm}) we obtain induced $\hat b^\pm$ potentials at first order
\be \hat b^\pm_{vac} = \frac{\theta}{ 2\ell^3}\;\; \frac
{\mp  dt
  \;+\; \frac { r^2}{8\ell} d\phi}{(1-\frac{ r^2}{4\ell^2})^2}\;. \ee
The resulting  connection one form $\hat  {\cal  A}^\pm=\hat  {\cal  A}^\pm_{vac} $ is
then just \be \hat {\cal   A}^\pm_{vac} =  {\cal  A}^\pm + \hat b^\pm_{vac} \tau_3+{\cal O}(\theta^2)
\;,\label{ncads3}\ee where $ {\cal  A}^\pm$ is the $ AdS^3$ vacuum solution in the
commutative theory,  which was expressable as a  pure gauge
(\ref{prgg}).  (\ref{ncads3}) satisfies the free noncommutative field
equations (\ref{nczceo}), which follows because it is also
 expressable as  a pure gauge  \be\hat {\cal   A}^\pm_{vac} =\hat g_{\pm}^{-1}\star
d \hat g_{\pm }\;, \label{ncpgptnl}\ee where, up to first order in $\theta$,
\be \hat g_{\pm} = \frac 1 { \sqrt{1-\frac
    { r^2}{ 4\ell^2}}}
\;\pmatrix{\biggl(1-\frac{\frac \theta{8\ell^2}}{1-\frac{
      r^2}{4\ell^2}}\biggr) \exp{ \frac{\pm it }{2\ell}}
& \pm  \frac{1 }{2\ell}\;\bar z\;\exp{ \frac{\pm it }{2\ell}}\cr \pm
\frac 1{2\ell}\;z\;\exp{ \frac{\mp i t
  }{2\ell}} & \biggl(1+\frac{\frac\theta{8\ell^2}}{1-\frac{ r^2}{4\ell^2}}\biggr)
\exp{ \frac{ \mp it }{2\ell^2}}\cr
}\;.\label{grplmntsstr} \ee (\ref{grplmntsstr}) represents the Seiberg-Witten
transformation on  gauge group elements.

\subsection{Noncommutative  conical space}

Next we  substitute the singular  solution (\ref{snglrptnl}) into (\ref{swm}).
This     induces nonvanishing $\hat b^\pm$ potentials at first
order,
\be \hat b^\pm_{sing} =\frac {\theta} 4 \biggl(m\pm \frac s\ell\biggr)^2\; \frac
{d\phi}{r^2} \;,\qquad r>>\sqrt{\theta}\;.\ee   That it goes like $\theta/r^2$ is
consistent with a dimensional argument.  Since the first order
correction is singular as $r\rightarrow 0$, the lowest order
expression for the Seiberg-Witten map
cannot be valid near the origin.  Again there are no corrections at this order  to the gravitational
fields, and
the resulting  connection one form $\hat  {\cal   A}^\pm $ is
then just \be \hat  {\cal  A}^\pm =\biggl(m\pm \frac s\ell\biggr)d\phi\; \tau_0 + \hat
b^\pm_{sing}\tau_3+{\cal O}(\theta^2)\;,\qquad r>>\sqrt{\theta}\;.\label{swmfsptnl} \ee   The $1/r^2$
factor in the  $\hat
b^\pm$ potentials implies that the free noncommutative field equations
are not satisfied, {\it  even after removing the origin }
 \be \hat  {\cal  F}^\pm \equiv d \hat  {\cal  A}^{\pm}
+ \hat  {\cal  A}^\pm\starwedge  \hat  {\cal  A}^\pm =-\frac {\theta \tau_3}{2 r^4}\; \biggl(m\pm
\frac s\ell\biggr)^2\;d^2x\;+\;{\cal O}(\theta^2)\;,\qquad r>>\sqrt{\theta}\;.\label{sngrncfs} \ee
 As a result the Seiberg-Witten map of the potential (\ref{snglrptnl}) induces a nonlocal source in the last field equation in (\ref{nczceo}).\footnote{On the other hand, the commutative field
 strength away from the origin is zero, and  if we Seiberg-Witten map that directly to the noncommutative field
 strength  we get zero.  Therefore for singular fields, the
 Seiberg-Witten map does not commute with derivations.}  Because of this
one cannot express the result as a pure gauge as in (\ref{ncpgptnl}), even with singular  $U(1,1)\times
 U(1,1)$ group elements.

 The commutative
conical solutions were obtained by performing a commutative gauge
transformation (\ref{ggtrns}) on (\ref{snglrptnl}).
Similarly, to obtain the noncommutative conic solutions we  perform a noncommutative gauge
transformation (\ref{ncggtrns}) on (\ref{swmfsptnl}) by $\hat g_\pm$ given in (\ref{grplmntsstr}).
To compute the first order noncommutative corrections we can use
\beqa & & \hat  {\cal  A}'^\pm\; =\; \hat g_\pm^{-1} \hat  {\cal  A}^\pm\hat g_\pm\; +\;\hat g_\pm^{-1}
d\hat  g_\pm \cr & &\\ & &\; +\; \frac
i2\theta^{\alpha\beta}\biggl(\partial_\alpha g_\pm^{-1}\partial_\beta   {\cal  A}^\pm g_\pm  +\partial_\alpha g_\pm^{-1}
 {\cal  A}^\pm \partial_\beta
g_\pm+ g_\pm^{-1}\partial_\alpha  {\cal  A}^\pm \partial_\beta g_\pm+ \partial_\alpha g_\pm^{-1}
d \partial_\beta g_\pm\biggr)+{\cal O}(\theta^2)\nonumber \eeqa
 Again there are no corrections at this order  to the gravitational
fields, and
the resulting  connection one form $\hat  {\cal  A}'^\pm $ is
then just \be \hat  {\cal  A}'^\pm =  {\cal  A}'^\pm + \hat b^\pm_{con} \tau_3+{\cal O}(\theta^2)
\;,\ee where $ {\cal  A}'^\pm$ is the conical solution in the
commutative theory, while  the $\hat b^\pm $ potentials are transformed to
\be \hat b^\pm_{con} =\hat b^\pm_{vac}  \;+\; \frac
 {\theta}{4\ell^2}\biggl(m\pm \frac s\ell\biggr)\;\;\biggl\{- \frac{ 1+\frac{
    r^2}{4\ell^2}}{(1-\frac{ r^2}{4\ell^2})^2} +\frac{\ell^2}{r^2}\biggl(m\pm \frac s\ell\biggr) \biggr\}\;d\phi\;,\qquad r>>\sqrt{\theta}\;.\ee
The result is also obtained by performing the Seiberg-Witten
transformation directly on the conical solution (\ref{scsln}).  This is due to the
fact that gauge transformations commute with the Seiberg-Witten map.\footnote{More generally, the commutator of the Seiberg-Witten map with
gauge transformations closes to gauge transformations.\cite{Grimstrup:2003rd},\cite{Pinzul:2004tq}}
Since
 the noncommutative field strength
(\ref{sngrncfs}) is in the $\tau_3$ direction and of order $\theta$ it
is unchanged at this order under a gauge transformation \be \hat
 {\cal  F}^\pm \rightarrow  \hat {\cal   F}'^\pm = \hat g_\pm^{-1}\star \hat
 {\cal  F}^\pm\star\hat g_\pm \;.  \ee

\subsection{$U(1)$ Fluxes}

Finally, we consider the general case where the sources also carry
$U(1)$ fluxes in the commutative theory.  Applying the lowest order
expression for the Seiberg-Witten
map to the potential $ {\cal  A}^\pm $ in  (\ref{snglrptnlwnzb}) gives
\be \hat  {\cal  A}^\pm = {\cal  A}^\pm   +
\frac {\theta} 4\biggl\{ \biggl(m\pm \frac s\ell\biggr)^2 \tau_3 +
2\Phi_\pm \biggl(m\pm \frac s\ell\biggr) \tau_0 +\Phi_\pm^2 \tau_3\biggr\} \frac
{d\phi}{r^2} \;+\;
{\cal O}(\theta^2)\;,\qquad r>>\sqrt{\theta} ,\ee  and so now there are
corrections at this order  to the triads and spin connections. As before the correction
breaks down near the origin. (\ref{sngrncfs}) is now replaced by \be \hat {\cal   F}^\pm \equiv d \hat {\cal   A}^{\pm}
+ \hat  {\cal  A}^\pm\starwedge  \hat  {\cal  A}^\pm = -
\frac {\theta} 2\biggl\{ \biggl(m\pm \frac s\ell\biggr)^2 \tau_3 +
2\Phi_\pm \biggl(m\pm \frac s\ell\biggr) \tau_0 +\Phi_\pm^2 \tau_3\biggr\} \frac
{d^2x}{r^4} \;+\;{\cal O}(\theta^2)\;,\qquad
r>>\sqrt{\theta}\label{sngrncfswflx}\;, \ee and
 nonlocal sources are now induced in all of  the
  field equations in (\ref{nczceo}).   Finally after performing the
  noncommutative
  gauge transformation (\ref{ncggtrns}), we get
\beqa \hat {\cal   A}_\pm' =  {\cal  A}'_\pm& +& \frac {\theta\Phi_\pm }
{2r^2}\;\frac  { \biggl(m\pm \frac
  s\ell\biggr)\biggl(1+\frac{r^2}{4\ell^2}\biggr)-\frac{r^2}{2\ell^2}}{1-\frac{ r^2}{4\ell^2}}\;  { d\phi}
\;\tau_0\cr & &\cr &\mp& \frac{\theta\Phi_\pm}{2\ell r^2}\;\frac {\hat x_i\hat
  x_j dx_j -\epsilon_{ij} x_j\biggl(m\pm \frac
  s\ell-1\biggr)d\phi}{1-\frac{ r^2}{4\ell^2}}\;\;\tau_i\\ & &\cr &+&
\frac{\theta}{4r^2}\Biggl\{\Phi_\pm^2+\biggl(m\pm \frac
  s\ell\biggr)^2 -\biggl(m\pm \frac
  s\ell\biggr) \frac{\frac{r^2}{\ell^2}\biggl(1+\frac{r^2}{4\ell^2}\biggr)}{\biggl(1-\frac{ r^2}{4\ell^2}\biggr)^2}  \Biggr\}\;  { d\phi}
\;\tau_3 \;+\;{\cal O}(\theta^2)\;,\qquad
r>>\sqrt{\theta} \nonumber \eeqa where $  {\cal  A}'_\pm$ is the commutative result.  Then
up to first order, the noncommutative
 triads and spin connections are
\beqa \hat e'^0 &=& \frac{1}{1-\frac{
    r^2}{4\ell^2}}\Biggl\{\biggr({1+\frac{
    r^2}{4\ell^2}}\biggl)\biggl[dt + \biggl(s +\frac{\theta
 }{4r^2}(m \ell\Phi_d + s\Phi_s) \biggr) d\phi\biggr]
-\frac{\theta\Phi_d}{8\ell} d\phi
\Biggr\} \cr & &\cr
\hat e'^i &=& \frac{1}{1-\frac{
    r^2}{4\ell^2}}\Biggl\{ \biggl(
m +\frac{\theta}{4\ell r^2}[(m-2)\ell \Phi_s + {s\Phi_d }]\biggr) \hat
x_i \hat x_ j dx_j\cr & &\cr & &\qquad\qquad\qquad \qquad\qquad
 -\biggl(m-1 +\frac{\theta}{4\ell r^2}[ (m-1)\ell \Phi_s + {s\Phi_d } ]\biggr)  dx_i\Biggr\}
\cr & &\cr
 \hat \omega'^0 &=& \frac{d\phi}{1-\frac{ r^2}{4\ell^2}}\Biggl\{
\biggr({1+\frac{ r^2}{4\ell^2}}\biggl)\biggl( m+ \frac\theta{4\ell r^2}
(m\ell\Phi_s + {s\Phi_d})\biggr)- \frac{ r^2}{2\ell^2}\biggl( 1 +
\frac{\theta\Phi_s}{4r^2}\biggr)\Biggr\} \cr & &\cr
\hat \omega'^i &=& \frac{\frac 1{\ell^2}}{1-\frac{
    r^2}{4\ell^2}}\Biggl\{\epsilon_{ij} x_j dt + \biggl(
s +\frac{\theta}{4r^2}[(m-2)\ell \Phi_d + {s\Phi_s }]\biggr) \hat
x_i \hat x_ j dx_j\cr & &\cr & &\qquad\qquad\qquad \qquad\qquad
 -\biggl(s +\frac{\theta}{4r^2}[ (m-1)\ell \Phi_d + {s\Phi_s } ]\biggr)  dx_i\Biggr\}
\;, \eeqa where $\Phi_s
=\Phi_+ + \Phi_-$ and  $\Phi_d=\Phi_+ - \Phi_-$.  In the limit of
large mass $m>>1$, and again for $r>>\sqrt{\theta}$, one can argue from the above result that there is
an effective $r-$dependent mass and spin
\beqa m_{eff}(r) &\approx &  m  + \frac\theta{4 r^2}
\biggl(m\Phi_s + \frac{s\Phi_d}\ell\biggr)\cr & &\cr
s_{eff}(r)&\approx & s\: +\:\frac{\theta}{4r^2}\;( {s\Phi_s }+ m\ell \Phi_d )\;.\eeqa

\subsection{The metric}

The physical interpretation of the solutions to the noncommutative
theory remain unclear without a noncommutative analogue of the metric
tensor.  The definition of the noncommutative metric is in general
ambiguous due to ordering problems.   A  definition in
\cite{Cacciatori:2002gq} which gives a real result for the case of the
Groenewold-Moyal product is the symmetric part of  \be \hat g_{\mu\nu} = \hat e_\mu^a \star
\hat e_\nu^b\; \eta_{ab}\;. \ee     To compute first order corrections from
it, we
can replace the star product by ordinary multiplication $\hat
g_{\mu\nu}+\hat g_{\nu\mu}= 2 \hat e_\mu^a
\hat e_\nu^b \eta_{ab}\;+\;{\cal O}(\theta^2)$, and as a result the $AdS^3$ vacuum  metric
(\ref{adsvcmmtrc}) receives no order $\theta$ corrections.  The same
is true for   the
 metric
(\ref{cmtvsnglrmtrc}) describing the conical space when no $U(1)$
fluxes are present.  If $U(1)$
fluxes are present the triads receive first order corrections, and
hence so does the metric
\beqa d\hat s^2 &=& d s^2+  \frac{\frac\theta{2r^2}}{(1-\frac{
    r^2}{4\ell^2} )^2}\;
\Biggl\{-\biggl(1+\frac{ r^2}{4\ell^2} \biggr)\biggl[\biggl(1+\frac{
  r^2}{4\ell^2} \biggr) (m \ell\Phi_d + s\Phi_s)-\frac{r^2
  \Phi_d}{2\ell} \biggr] (dt + s \;d\phi)d\phi    \cr & &\cr & &\qquad\qquad\;-\;
\Phi_s dr^2    \:+\: (m-1) \biggl[ (m-1) \Phi_s +\frac s\ell
\Phi_d\biggr] r^2 d\phi^2          \Biggr\}\;+\;{\cal O}(\theta^2)\;,\qquad
r>>\sqrt{\theta}\;,\nonumber\eeqa where $ds^2$ is
given by (\ref{cmtvsnglrmtrc}).  Similarly, in \cite{Banados:2001xw} it was found that
the only  nontrivial first order noncommutative  corrections to the
BTZ black hole solution occur when the $U(1)$ fluxes are turned on.

\section{Concluding remarks}

In this article we argued that the cosmological constant (more precisely,
$\Lambda G^2$) gets quantized in noncommutative three-dimensional
Chern-Simons gravity based on $U(1,1)\times U(1,1)$.   It would be of
interest to explore whether this result can be  generalized to higher
dimensions.

 We also  mapped solutions of commutative Chern-Simons gravity
to the corresponding  noncommutative theory, and   computed the lowest
order corrections to  the metric and the $U(1)$ fields. However,
the physical interpretation of
the noncommutative theory remains problematic.  An alternative approach would be to search for
solutions of the noncommutative theory and then map back to the
commutative theory where the interpretation is that  of standard  Einstein gravity
(at least when the metric is nonsingular).
Both of these approaches however  are limited by the fact that
the Seiberg-Witten map is in general only known perturbatively.
Moreover, for conical spaces, we see that the perturbative expansion breaks down near the singularity
$r\sim\sqrt{\theta}$.  It is often believed that the singularities which occur in
general relativity can be removed by passing to a noncommutative
version of the theory.  Unfortunately here, since  the Seiberg-Witten map
breaks down near the origin  one cannot  check whether this is the
case for the conical singularity, at least the perturbative level, and  the prospects
of a nonperturbative solution seems difficult.  Clarification of this
issue should come upon replacing the point source in the commutative theory with a nonlocal
distribution
arising from some  matter fields.  A Seiberg-Witten map can then be
performed directly on  the matter fields which would induce a nonlocal
source for the $U(1,1)\times
 U(1,1)$ gauge fields, replacing the obscure right hand side of
 (\ref{sngrncfs}).

   Yet another approach to understanding the region near the conical
 singularity in the noncommutative theory
 is to drop the restriction of a constant noncommutative
structure (\ref{cmtrxoxt}).  One should then also adopt a more general star product,
such as the one developed in \cite{Alexanian:2000uz} based on
 nonlinear coherent states.  This approach
has been taken to analyze the noncommutative analogue of  the punctured plane
in \cite{Pinzul:2001my}.  There it was found that near the commutative
 limit, and far from the puncture,
$\theta$ was constant, hence corresponding to
 the noncommutative plane.  $\theta$ went linearly to
zero as the puncture was approached, which may lead to a more
 consistent description of  noncommutative conical spaces.  In the context of $U(1)$ Chern-Simons theory,
degrees of freedom associated with a deformed $w_\infty$ algebra
were found to be localized near the puncture for small
$\theta$.\cite{Pinzul:2002fi}  Similar  results are then expected for
the $U(1,1)\times
 U(1,1)$ Chern-Simons theory  studied here, with gravitational degrees
 of freedom located in the
 vicinity of the conical singularity.

\end{document}